\documentclass[twocolumn,a4paper,showpacs,prl,aps]{revtex4-1}

\usepackage[dvipsnames]{xcolor}
 \usepackage{amsmath}
 \usepackage{amssymb}
 \usepackage{bbold}
 \usepackage{latexsym}
 \usepackage{amsfonts}
 \usepackage[caption=false]{subfig}
 \usepackage{epsfig}
 \usepackage{psfrag}
 \usepackage{color}
 \definecolor{darkblue}{rgb}{0,0,.5}
 \usepackage[linktocpage, colorlinks=true ,linkcolor=darkblue, citecolor=darkblue]{hyperref}
 \usepackage[all]{hypcap}
 
\usepackage{graphicx}
\usepackage{dcolumn}
\usepackage{bbm}
\usepackage{bm}
\usepackage{overpic}
  \newcommand{\ket}[1]{\left|#1\right>}
  \newcommand{\bra}[1]{\left<#1\right|}
  \newcommand{\expval}[1]{\left< #1 \right>}

\begin{document}

\title{Quantum criticality and dynamical instability in the kicked top}
%%%%%%%%%%%%%%%%%%%%%%%%%%%%%%%%%%%%%%%%%%%%%%%%%%%%%%%%%%%%%%%%

\author{Victor Manuel Bastidas$^1$}\email{victor@physik.tu-berlin.de}
\author{Pedro P{\'e}rez-Fern{\'a}ndez$^{1,2}$}\email{pedropf@us.es}
\author{Malte Vogl$^{1}$}\email{mvogl@pks.mpg.de}\altaffiliation[Current address: ]{Max Planck Institute for the Physics of Complex Systems, N\"othnitzer Stra\ss e 38, 01187 Dresden, Germany}
\author{Tobias Brandes$^1$}

\affiliation{$^1$Institut f\"ur Theoretische Physik, Technische Universit\"at 
Berlin, Hardenbergstr. 36, 10623 Berlin, Germany}
\affiliation{$^2$Departamento de F\'isica Aplicada III, Escuela Superior de Ingenieros, 
Universidad de Sevilla, Camino de los Descubrimientos s/n, ES-41092 Sevilla,
Spain}

%%%%%%%%%%%%%%%%%%%%%
\begin{abstract}
%%%%%%%%%%%%%%%%%%%%%

We investigate precursors of critical behavior in the quasienergy spectrum due to the dynamical instability in the kicked top. Using a semiclassical approach, we analytically obtain a logarithmic divergence in the density of states, which is analogous to a continuous excited state quantum phase transition in undriven systems. We propose a protocol to observe the cusp behavior of the magnetization close to the critical quasienergy.
%%%%%%%%%%%%%%%%%%%
\end{abstract}
%%%%%%%%%%%%%%%%%%%


\maketitle
%%%%%%%%%%%%%%%%%%%%%%%%%%%%%%%%%%%%%%%%%%%%%%%%%%%%%%%%%%%%%%%%%%%%%%%%%%%%%%%%
%%%%%%%%%%%%%%%%%%%%%%%%%%%%%%%%%%%%%%%%%%%%%%%%%%%%%%%%%%%%%%%%%%%%%%%%%%%%%%%%
The emerging field of excited state quantum phase transitions (ESQPTs) 
describes the nonanalytical behavior of excited states upon changes of 
parameters in the Hamiltonian~\cite{CejnarA,Caprio08,CejnarB}. 
This is in direct correspondence to quantum phase transitions (QPTs)~\cite{Sachdev}, but takes place 
at critical energies above the ground state energy~\cite{Leyvraz2005}.
They entail dramatic dynamic consequences, e.g.,  environments with ESQPTs lead to enhanced decoherence, which could be a major drawback for building a quantum computer~\cite{PPF08A}. 
They appear in models of nuclear physics, such as the interacting 
Boson model~\cite{Arias03, CejnarRMF} and the 
Lipkin-Meshkov-Glick model (LMG)~\cite{RibeiroESQPT}.
In molecular physics, singularities of the density of states (DOS) arise in the vibron model~\cite{Curro10},
which are closely related to the monodromy in molecular bending degrees of 
freedom~\cite{monodromy}.
ESQPTs have been predicted to occur in prominent models of quantum optics such 
as the Dicke and Jaynes-Cummings models~\cite{BrandesESQPT,PPF11E}, too. 

Despite the striking observation of the QPT in the Dicke and the LMG model~\cite{Esslinger,Oberthaler}, 
ESQPTs have so far not been found experimentally for systems different to molecular ones, as 
the energies at which they occur are difficult to reach with standard 
techniques. 
Recently, however, the observation of low-energy singularities of the DOS in 
twisted graphene layers~\cite{Andrei}, and monodromy in diverse molecules~\cite{monodromy},
has opened an increasing interest in the experimental investigation of spectral singularities. 

Quantum critical behavior is usually defined with respect to system energies~\cite{Sachdev}. 
Under the effect of a nonadiabatic external control the energy is not conserved and it is not possible to uniquely define a ground state and the corresponding excited states. 
In this paper we make use of Floquet theory to introduce the concept of critical quasienergy states (CQS), which are a direct generalization of ESQPTs to driven quantum systems.
Our model of choice is a paradigmatic model in the quantum chaos community: the kicked top.
Quantum kicked systems play a prominent role in the investigation of quantum signatures of chaos and have intriguing relations to condensed matter systems~\cite{HaakeBuch}. 
Examples of these relations are the metal-supermetal~\cite{Altland} and metal-topological-insulator~\cite{Dahlhaus1} QPTs in the kicked rotator,
which can be thought of as a limiting case of the kicked top.
Such a limit is established when the top is restricted to evolve along a small equatorial band, which is topologically equivalent to a cylinder~\cite{HaakeBuch}.

We are motivated by a recent experimental realization of the 
kicked top with driven ultra-cold Cesium-atoms~\cite{Chaudhury1,Chaudhury2}.
The large body of previous work on the kicked top is mostly focused on aspects 
of chaotic dynamics~\cite{Haake1,Haake2,Haake3, Haake4, Haake5}. 
Contrary to this, our approach is strictly limited to the 
regular regime, taking advantage of the well-known level clustering~\cite{HaakeBuch}.  

We show that in this regime, the kicked top can be described 
by means of a time-independent effective Hamiltonian, which allows one to study the dynamical instabilities leading to CQS.
Signatures of CQS then appear in the density of quasienergy states (DOQS), and the transverse magnetization.

%%%%%%%%%%%%%%%%%%%%%%%%%%%%%%%%%%%%%%%%%%%%%%%%%%%%%%%%%%%%%%%%%%%%%%%%%%%%%%%%%%%%%%%%%%%%%%%%%%%%%%%%%%%%%%%%%%%%%%%%%%%
\paragraph{Model and method.---}
We consider the Hamiltonian for the kicked top~\cite{Haake1,Haake2}
%%%
\begin{equation}
      \label{KickedTopHamiltonian}
            \hat{H}(t)=pJ_{x}\sum_{n=-\infty}^{\infty}\delta(t-nT)+\frac{\kappa}{2jT}J_{z}^{2}\ ,
\end{equation}
%%%
where $J_{\nu}$ with $\nu \in \{x,y,z,\pm\}$ denote collective angular momentum operators with   
total angular momentum  $j$. 
The time dependent term in the Hamiltonian~\eqref{KickedTopHamiltonian} describes a rotation along the $x$-axis, which acts stroboscopically with strength $p$. 
The second term describes a twisting with strength $\kappa$, i.e., a rotation where the angle depends on the state of the system~\cite{Haake1,Haake2}.
%%%%
%%%%
\begin{figure}
\hspace{-0.65cm}
  \begin{minipage}[t]{\linewidth}%grid,tics=5,
    \begin{overpic}[clip=true,width=\linewidth]{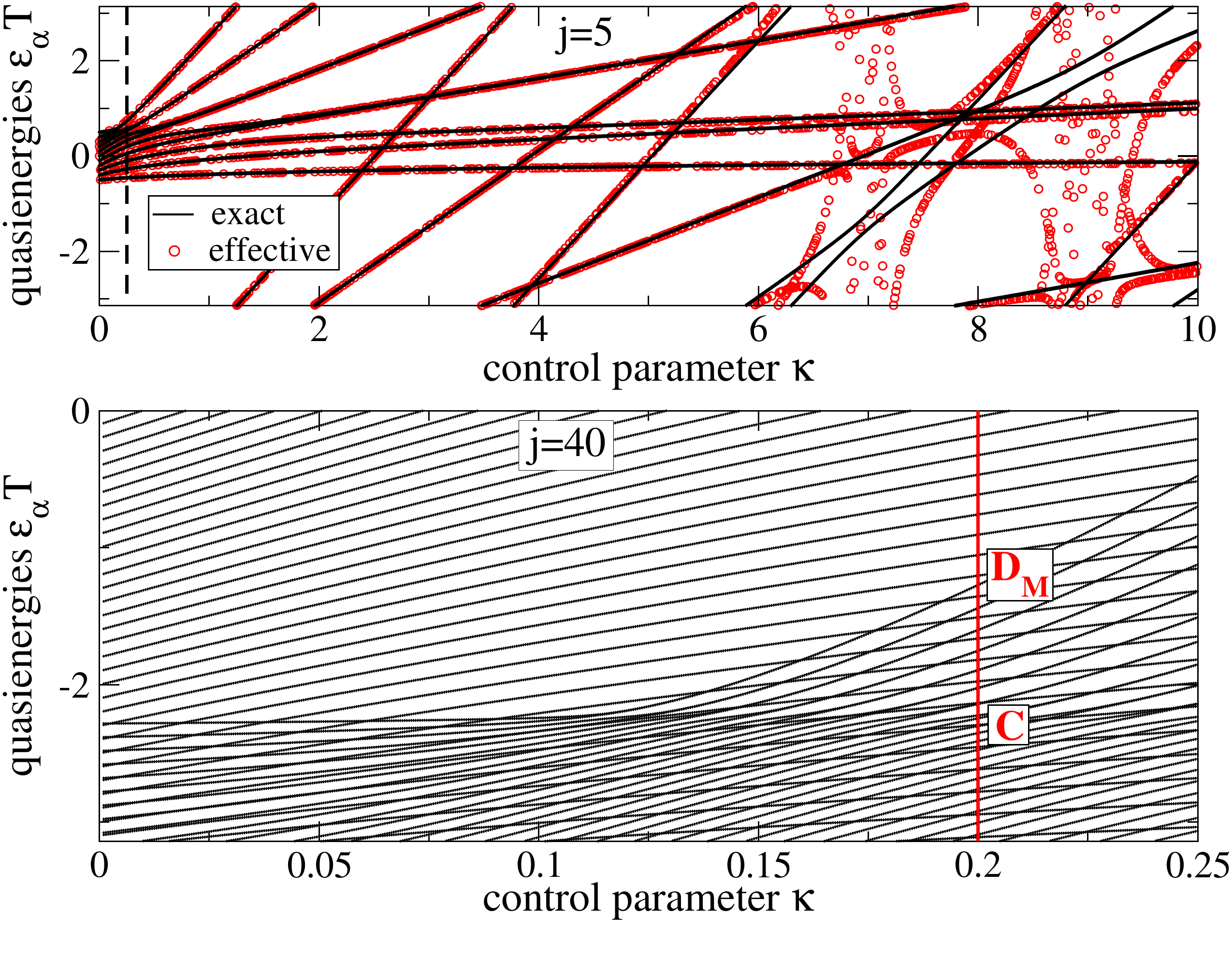}
      \put(2,80){\textbf a)}
      \put(2,47){\textbf b)}
    \end{overpic}
  \end{minipage}
\caption{(color online).
(a) Exact quasienergy spectrum (black lines) and the eigenvalues of 
the effective Hamiltonian (red, open circles) as a function of $\kappa $ for $j=5$. 
(b) Zoom into the regime $\kappa \ll 1$ 
[dashed vertical line in Fig.~\ref{Fig1} a)] 
for $j=40$. 
The red, vertical line denotes the value of $\kappa$ used in Figure~\ref{Fig2}. 
Features are marked by C (clustering) and $D_M$ (discontinuity), see main text.
The parameters are $p=0.1$ and $T=1$.}
    \label{Fig1}
\end{figure}
%%%%
%%%%

As the Hamiltonian~\eqref{KickedTopHamiltonian} is periodic in time, it is convenient to use Floquet theory~\cite{Shirley,Sambe,HanggiGrifoni}. 
The Floquet operator $\hat{\mathcal{F}}=\hat{U}(T)$ is defined as the evolution operator after one period $T=2\pi/\omega$ of the external driving.
The Floquet modes $\ket{\Phi_{\alpha}(t)}= \ket{\Phi_{\alpha}(t+T)}$ satisfy the eigenvalue problem $\hat{\mathcal{F}}\ket{\Phi_{\alpha}(0)} =e^{-\mathrm{i} \varepsilon_{\alpha} T} \ket{\Phi_{\alpha}(0)}$, where $\varepsilon_{\alpha}$ are the quasienergies. 
These are not unique as there are equivalent states $\ket{\Phi_{\alpha,n}(0)}=
e^{\mathrm{i} n\omega T}\ket{\Phi_{\alpha}(0)}$, such that $\hat{\mathcal{F}}\ket{\Phi_{\alpha,n}(0)} =e^{-\mathrm{i} \varepsilon_{\alpha,n} T} \ket{\Phi_{\alpha,n}(0)}$, with $\varepsilon_{\alpha,n}=\varepsilon_{\alpha}+n\omega$ and $n \in \mathbb{Z}$. 
In what follows, we consider the quasienergies 
in the first Brillouin zone defined by $- \omega/2\leq \varepsilon_{\alpha} \leq \omega/2$~\cite{Shirley,Sambe,HanggiGrifoni}. Without loss of generality we choose $T=1$ throughout the paper.

Since within one period the system
evolves under free evolution followed by an impulsive force, the Floquet 
operator for the Hamiltonian~\eqref{KickedTopHamiltonian} reads
%%%
\begin{equation}
      \label{FloqOpKick}
	    \hat{\mathcal{F}}=e^{-\mathrm{i} pJ_{x}} e^{-\mathrm{i} (\kappa/2j) J^{2}_{z} }
      \ .
\end{equation}
%%%
Motivated by Eq.~\eqref{FloqOpKick}, we construct an effective Hamiltonian $\hat{H}_{\text{E}}$ such that $\hat{\mathcal{F}}=e^{-\mathrm{i}\hat{H}_{\text{E}}}=e^{-p\hat{B}}e^{-\hat{A}}$, with $\hat{A}=\mathrm{i}\frac{\kappa}{2j}J^{2}_{z}$ and $\hat{B}=\mathrm{i} J_{x}$. 
We use the Baker-Campbell-Hausdorff (BCH) formula in the regime $p\ll1$ to construct  $\hat{H}_{\text{E}} \approx  -\mathrm{i}\hat{A}+\mathrm{i} p \frac{\text{ad}_{\hat{A}}}{\exp[-\text{ad}_{\hat{A}}]-1}\hat{B}$, where $\text{ad}_{\hat{X}}\hat{Y}=[\hat{X},\hat{Y}]$ is the adjoint representation of the angular momentum algebra~\cite{Scharf1}.
One can analytically calculate the $n$th power of the adjoint action of $\hat{A}$ on $\hat{B}$
%%%
\begin{eqnarray}
      \label{PowerAdjoint}   
            (\text{ad}_{\hat{A}})^{n}\hat{B} = \frac{1}{2}\left(\mathrm{i}\frac{\kappa}{2j}\right)^{n}\left[J_{+}(2J_z+1)\right]^{n}+\text{H.c}
      \ ,
\end{eqnarray}
%%%
which allows us to construct
%%%
\begin{eqnarray}
      \label{EffectiveHamiltonianKicked}   
            \hat{H}_{\text{E}}
             = \frac{\kappa}{2j}J^{2}_{z}+\frac{p}{2}\biggl[\frac{-\mathrm{i}\frac{\kappa}{2j}J_{+}(2 J_{z}+1)}{\exp(-\mathrm{i}\frac{\kappa}{2j}(2 J_{z}+1))-1}+\text{H.c}\biggr]            
\end{eqnarray}
%%%
by using $\frac{z}{e^z-1}=\sum_{n=0}^{\infty}B_n\frac{ z^n}{n!}$, where $B_n$ are the Bernoulli numbers~\cite{Scharf1}.
The Floquet modes $\ket{\Phi_{\alpha}(0)}$ are eigenvectors of~\eqref{EffectiveHamiltonianKicked},
$\hat{H}_E \ket{\Phi_{\alpha}(0)}=E_\alpha \ket{\Phi_{\alpha}(0)}$. 
Correspondingly, the eigenvalues $\{E_\alpha\}$ of $\hat{H}_\text{E}$ are \textit{unfolded} quasienergies, which approximate the genuine quasienergies $\{\varepsilon_\alpha\}$ when they are mapped into the first Brillouin zone, i.e., $\varepsilon_\alpha = E_\alpha \ \text{mod} \  \omega$.
By using $\hat{H}_\text{E}$ we approximate the non-integrable
Floquet operator~\eqref{FloqOpKick} by an integrable one~\cite{HaakeBuch}, leading to crossings in the spectrum of $\hat{H}_\text{E}$ after folding. Our exact numerical results show that for states in one symmetry multiplet, the apparent crossings in the quasienergy spectrum are instead anticrossings~\cite{Scharf1}. In the regime $p\sim \kappa \ll 1$, the anticrossings are neglectable compared with the mean level spacing and the effective Hamiltonian is a good approximation to describe the stroboscopic dynamics~\cite{Scharf1}.
To stay in a valid parameter regime we will consider a kick-strength of $p=0.1$ and $\kappa \ll 1$ for the remainder of the paper. 
Figure~\ref{Fig1} a) shows the excellent agreement between the eigenvalues of $\hat{H}_\text{E}$ and the exact quasienergy spectrum.
However, when the system approaches the chaotic regime for $\kappa \gg 1$, the approximation is not valid anymore. 
In a basis such that $J_{z}\ket{j,m}=m\ket{j,m}$, $\hat{H}_\text{E}$ then has singularities in the nondiagonal matrix elements at values $\kappa(2m+1)=4jl\pi$, for $l \in {\mathbb Z}$.

%%%%%%%%%
\paragraph{Spectral signatures of the CQS .---}
%%%%%%%%%%

For undriven models the hallmark of a second order ESQPT is the existence of logarithmic divergences in the DOS, which arise due to saddle points of semiclassical energy landscapes~\cite{Caprio08,RibeiroESQPT,BrandesESQPT,CejnarRMF}.
As a first test whether similar signatures can be found for driven models, Fig.~\ref{Fig1} b) shows the quasienergy spectrum obtained numerically from Hamiltonian~\eqref{KickedTopHamiltonian} in the regular regime. 
For $\kappa > p=0.1$, a level clustering, marked by $C$, is visible around the critical quasienergy $\varepsilon_{S} T=jp \ \text{mod} \  \omega \approx -2.283$. 

In addition, a discontinuity of the DOQS is marked by $D_M$, which arise due the confinement of the quasienergies to the first Brillouin zone.

In the regular regime $p\sim\kappa\ll 1$, the semiclassical version of Hamiltonian~\eqref{KickedTopHamiltonian} exhibits a dynamical instablity~\cite{HaakeBuch,Haake1,Haake2}. This occurs when a stable fixed point for $\kappa<p$ becomes unstable at $\kappa=p$, and two new stable fixed points arise for $\kappa>p$. The level clustering in Fig.~\ref{Fig1} b) is a quantum signature of the dynamical instablity, and gives rise to the CQS with critical quasienergy $\varepsilon_{S}$.

In the following we derive analytic expressions for the DOQS, the logarithmic divergence and the magnitude of the discontinuity. 
%%%%%%%%%%%%%
\paragraph{The density of quasienergy states.---}
%%%%%%%%%%% 
Following a method similar to Ref.~\cite{Haake2}, one obtains the 
DOQS in terms of the traces $t_{n}=\text{tr}\hat{\mathcal{F}}^{n}$ of the 
Floquet operator $\hat{\mathcal{F}}$ in Eq.~\eqref{FloqOpKick}, 
%%%
\begin{equation}
      \label{DefDOQS}
            \rho(\varepsilon)=\frac{1}{2\pi}+\frac{1}{\pi(2j+1)} {\rm Re}\left[\sum_{n=1}^{\infty}t_{n}e^{\mathrm{i} n \varepsilon }\right]
       \ .
\end{equation}
%%%
See supplemental informations online for some intermediate steps of the full derivation~\cite{SuppInfo}.

By using spin coherent states $\ket{\gamma}$~\cite{NoriSpinSqueez}, the traces are given by an integral over the Bloch sphere
%%%
\begin{equation}
      \label{SemiclassTraces}
            t_{n}= \frac{2j+1}{\pi}\int \frac{d^{2}\gamma}{(1+\gamma\gamma^{\ast})^2}e^{-\mathrm{i} n jE_{G}(\gamma,\gamma^{\ast})}
       \ ,
\end{equation}
%%%
where $E_{G}(\gamma,\gamma^{\ast}) \equiv \frac{1}{j}\bra{\gamma}\hat{H}_\text{E}\ket{\gamma}$ is the semiclassical quasienergy landscape (QEL), which allows to describe the dynamical instability.

For a large total angular momentum $j\gg 1$, the QEL in Cartesian coordinates is obtained by replacing 
the commuting variables $\textbf{R}=\left(X,Y,Z\right)=\left(J_{x}/j,J_{y}/j,J_{z}/j\right)$ in Eq.~\eqref{EffectiveHamiltonianKicked}, as follows
 %%%
 \begin{equation}
       \label{QELKicked}
             E_{G}(\textbf{R})=\frac{\hat{H}_{E}}{j}=\frac{\kappa}{2}Z^2+\frac{\kappa p Z}{2}\biggl[X\cot\left(\frac{\kappa Z}{2}\right)-Y \biggr]\,.
 \end{equation}
 %%%
A transformation to stereographic coordinates $\gamma=u+\mathrm{i}v$ can be obtained via $(X,Y,Z)= \left(\frac{1-\gamma\gamma^{\ast}}{1+\gamma\gamma^{\ast}},\frac{2{\rm Im}\gamma}{1+\gamma\gamma^{\ast}},\frac{-2{\rm Re}\gamma}{1+\gamma\gamma^{\ast}}\right)$.
In these coordinates, we denote the critical 
points $\textbf{R}_{c}$ of the QEL by $\gamma_{c}=u_{c}+\mathrm{i}v_{c}$, where $c$ denotes the different critical points. 
In the regime $\kappa<p$, the QEL exhibits a minimum $m$ and a maximum $M$ ($c \in \mathcal{C}_{1}=\{m,M\}$), while for $\kappa>p$ the QEL exhibits a minimum $m$, a saddle point $S$ and two maxima $M_{1}$ and $M_{2}$ 
($c \in \mathcal{C}_{2}= \{m, S, M_{1},M_{2}\}$).
Similarly, we define unfolded critical quasienergies associated with the critical points as $E_{c}=jE_{G}(\textbf{R}_{c})$. 

In the limit $j\gg 1$, we invoke the stationary-phase approximation~\cite{Haake2,HaakeBuch} to calculate the integral Eq.~\eqref{SemiclassTraces}.
The DOQS is thus a sum over the critical points
%%%
\begin{equation}
      \label{QEDensity}
            \rho(\varepsilon)=\frac{1}{2\pi}+{\rm Re}\left \{\sum_{c \in \mathcal{C}_{1,2}} A_{c} \ e^{\mathrm{i}\beta_{c}\pi/4} {\rm Li}_{1}\left[e^{\mathrm{i}(\varepsilon-E_{c})}\right] \right\}
      \ ,
\end{equation}
%%%
where $\text{Li}_{r}(z)=\sum^{\infty}_{n=1}\frac{z^{n}}{n^{r}}$ is the 
polylogarithm~\cite{Abramowitz}. 
The index $\beta_{c}$ takes the 
values $\beta_{M}=-\beta_{m}=2$ at the maxima and minima respectively, and 
$\beta_{S}=0$ for the saddle point. 
The amplitudes for each critical point 
$A_{c}$ are given by
%%%
\begin{equation}
      \label{Amplitude}
            A_{c}= \frac{2(1+\gamma_{c}\gamma^{\ast}_{c})^{-2}}{\pi j \sqrt{|\det\left[\text{HE}_{G}(\gamma,\gamma^{\ast})\right]|_{\gamma=\gamma_{c}}}}
      \ ,
\end{equation}
%%%
where $\text{HE}_{G}(\gamma,\gamma^{\ast})$ is the Hessian matrix of the QEL of Eq.~\eqref{QELKicked}. 
%%%%
%%%%
\begin{figure}
    \includegraphics[clip=true,width=\linewidth]{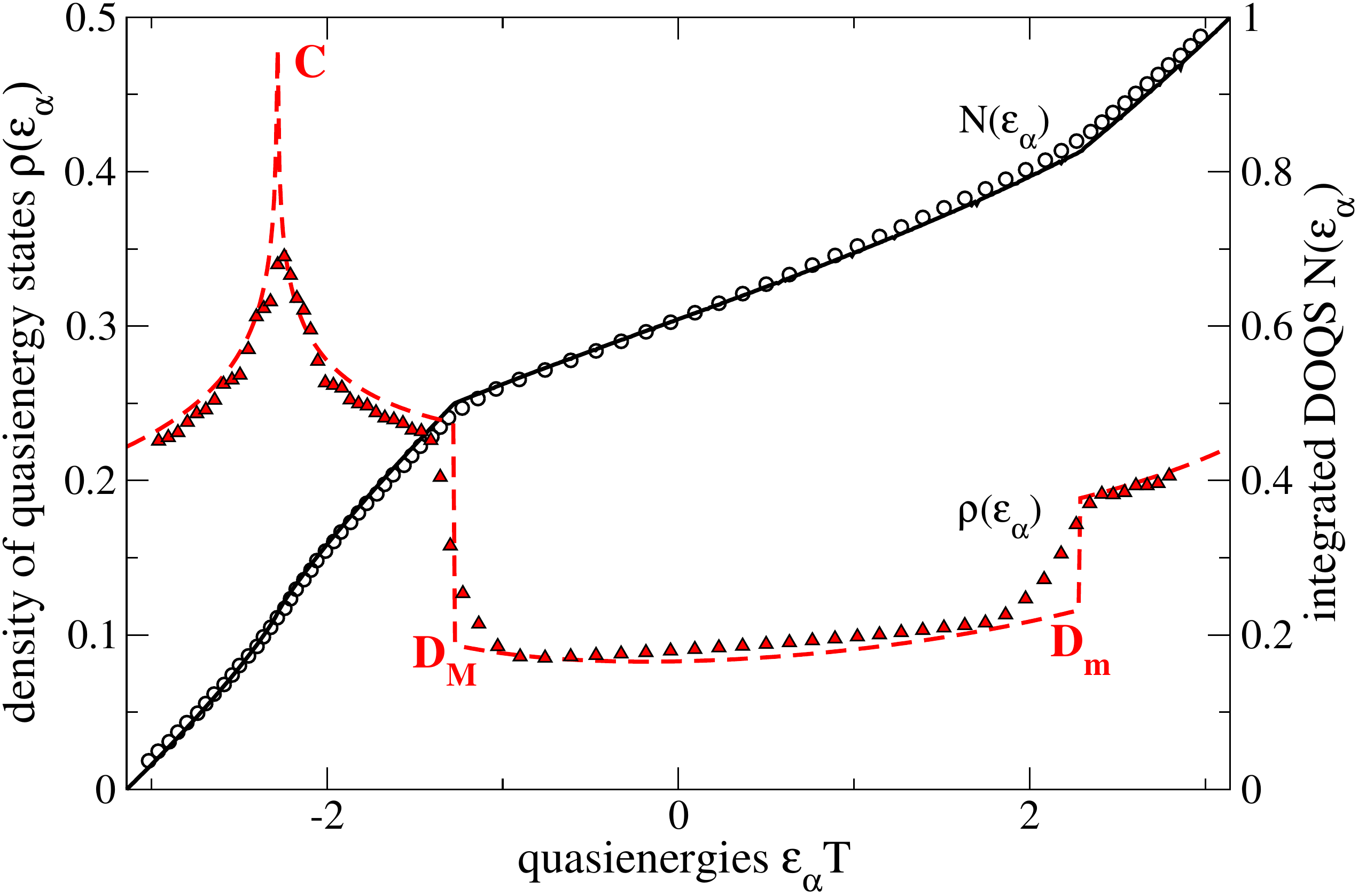}
    \caption{(color online). Comparison between the analytical (solid and dashed curves) and exact numerical
results (symbols) for the DOQS (dashed, red) and the integrated DOQS  (solid, black). The behavior is in 
direct correspondence with the spectrum along the vertical solid line in Fig.~\ref{Fig1} (b).
Features are marked by 
$C$ (clustering) and $D_{M,m}$ (discontinuity due to (M)axima and (m)inimum).
    The parameters are $j=40$, $\kappa=0.2$, $p=0.1$, and $T=1$.}
    \label{Fig2}
\end{figure}
%%%%
%%%%

In order to extract the singularities in the DOQS in the kicked top, we consider the expression for the polylogarithm~\cite{Abramowitz}
%%%
\begin{equation}
      \label{PolyLog}
            {\rm Li}_{1}\left(e^{\mathrm{i}\theta}\right)
            =-\log\left[2\sin\left(\frac{\theta}{2}\right)\right]+\mathrm{i}\left(\frac{\pi-\theta}{2}\right)
     ,
\end{equation}
%%%
where $0\leq\theta<2\pi$. 
The exponential function in the argument of the polylogarithm in Eq.~\eqref{QEDensity} maps the unfolded quasienergies automatically into the first Brillouin zone. 
Therefore, in the discussion of the singularities of the DOQS, we refer to genuine quasienergies.

In the saddle point the index $\beta_{S}$ vanishes, such that close to the critical quasienergy  $\varepsilon_{S}$, we extract the nontrivial behavior of the DOQS by using the real part of the expansion of  Eq.~\eqref{PolyLog} for small $\theta$
%%%
\begin{equation}
     \label{LogDivergence}         
            \rho(\varepsilon) \approx -A_S\log|\varepsilon-\varepsilon_{S}|
      \ ,
\end{equation}
%%%
which indeed reveals a logarithmic divergence. 
For the parameters of Fig.~\ref{Fig2}, $A_S\approx 0.0396$, which is consistent with a fit of the numerical result.
The effect of the minima and maxima is to create jumps in the density of states. 
To study these, one has to take into account that $\beta_{M}=-\beta_{m}=2$, which introduces an imaginary unit in Eq.~\eqref{QEDensity} for each maximum or minimum.
Based on this, we calculate the discontinuity at $\varepsilon_{m,M}$ by considering the imaginary part of Eq.~\eqref{PolyLog}
%%%
\begin{equation}
      \label{DOQSJumps}
            \rho(\varepsilon^{+}_{m,M})-\rho(\varepsilon^{-}_{m,M})= \pm \pi 
            A_{m,M}+\mathcal{O}[\varepsilon-\varepsilon_{m,M}]
      \ ,
\end{equation}
%%%
where $\varepsilon^{\pm}_{m,M}$ denote the limits from the right and from the left of the critical quasienergies $\varepsilon_{m,M}=E_{m,M}\ \text{mod} \  \omega$, respectively.

In the regime $\kappa < p$ there is no divergence of the DOQS as the QEL does 
not exhibit saddle points and therefore, there is no dynamical instability.
However, the maxima and minima still
generate jumps in the DOQS at $\varepsilon_{m,M}$. 
In undriven models these jumps would indicate a first order ESQPT~\cite{CejnarB}. In our driven model, however, they are just a consequence of the periodicity of the quasienergies.

Figure~\ref{Fig2} shows 
the comparison between the analytical result of Eq.~\eqref{QEDensity} and exact 
numerical calculations of the DOQS in the regime 
$\kappa>p$. 
The clustering at $\varepsilon_S T \approx -2.283$ is marked by C. 
The discontinuities due to the maxima ($D_M$) and minimum ($D_m$) of the QEL at $\varepsilon_M T \approx -1.288$ and $\varepsilon_m T \approx 2.283$, respectively, have
corresponding amplitudes of $A_M = 2 A_m \approx 0.046$.
Even for a finite size $j=40$, the system exhibits precursors of the 
logarithmic divergence at $\varepsilon_{S}$. 
To support our findings, we have 
incorporated in Fig.~\ref{Fig2} the integrated DOQS 
$N(\varepsilon)$, where the features of the DOQS are visible as 
a discontinuous change of slope at the critical quasienergies.

%%%%%%%%%%%%%
\paragraph{The transverse magnetization.---}
%%%%%%%%%%%

The CQE are the natural generalization of ESQPTs to driven quantum systems.
In particular, the CQE in the kicked top also appears in observables of the system, as they inherit the singular behavior of the DOQS~\cite{RibeiroESQPT, BrandesESQPT}
This is a direct consequence of the extension of the Hellmann-Feynman theorem to Floquet theory~\cite{HanggiGrifoni}.
For convenience, in the description of the dependence of observables with respect to the quasienergies, we restrict ourselves to the unfolded quasienergies, as they have a natural ordering and resemble the behavior of an effective time independent system.

\begin{figure}    
    \begin{minipage}[t]{\linewidth}
    \begin{overpic}[clip=true,width=\linewidth]{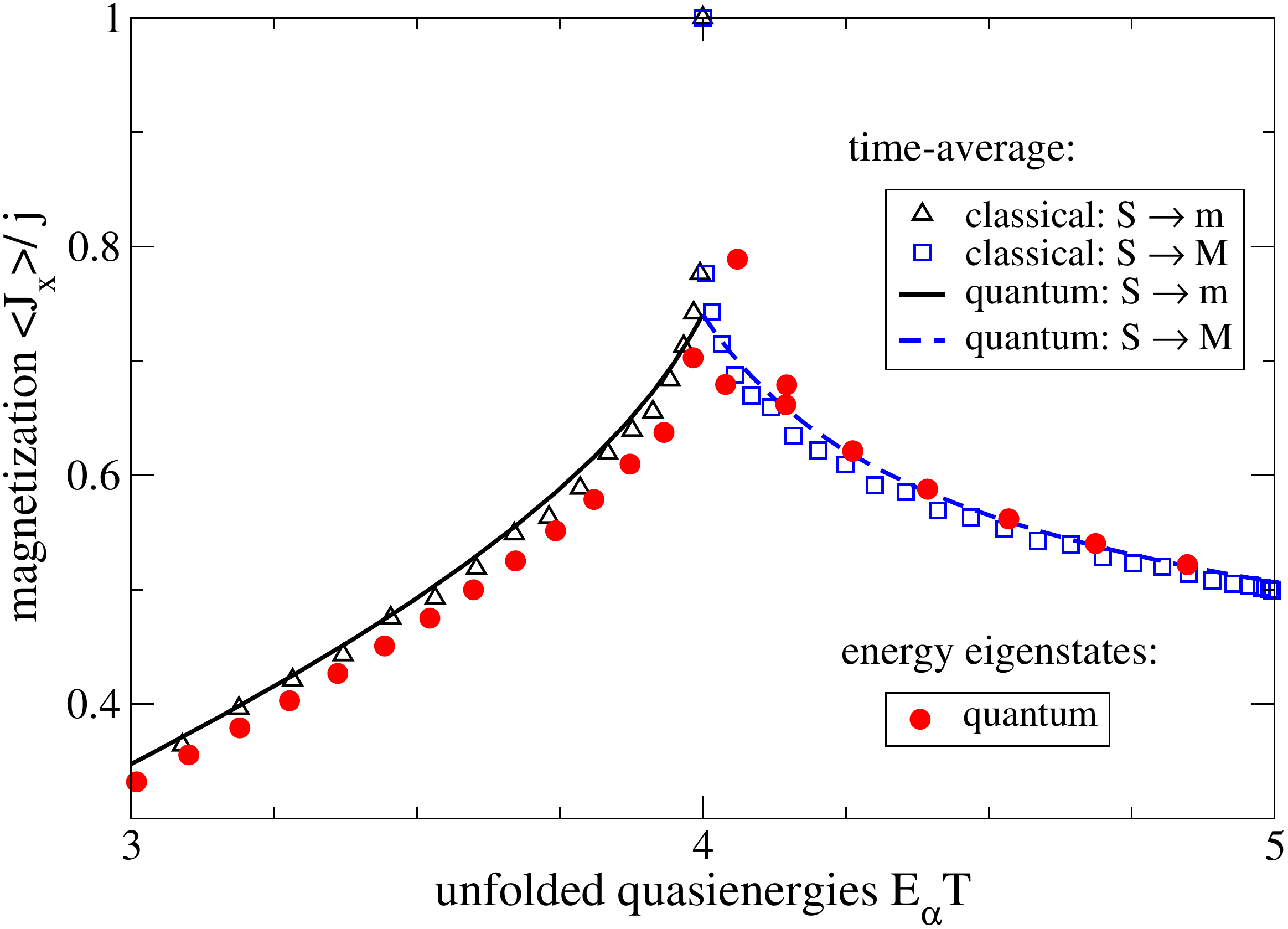}
	\put(12,37){\includegraphics[width=0.32\linewidth,clip=true]{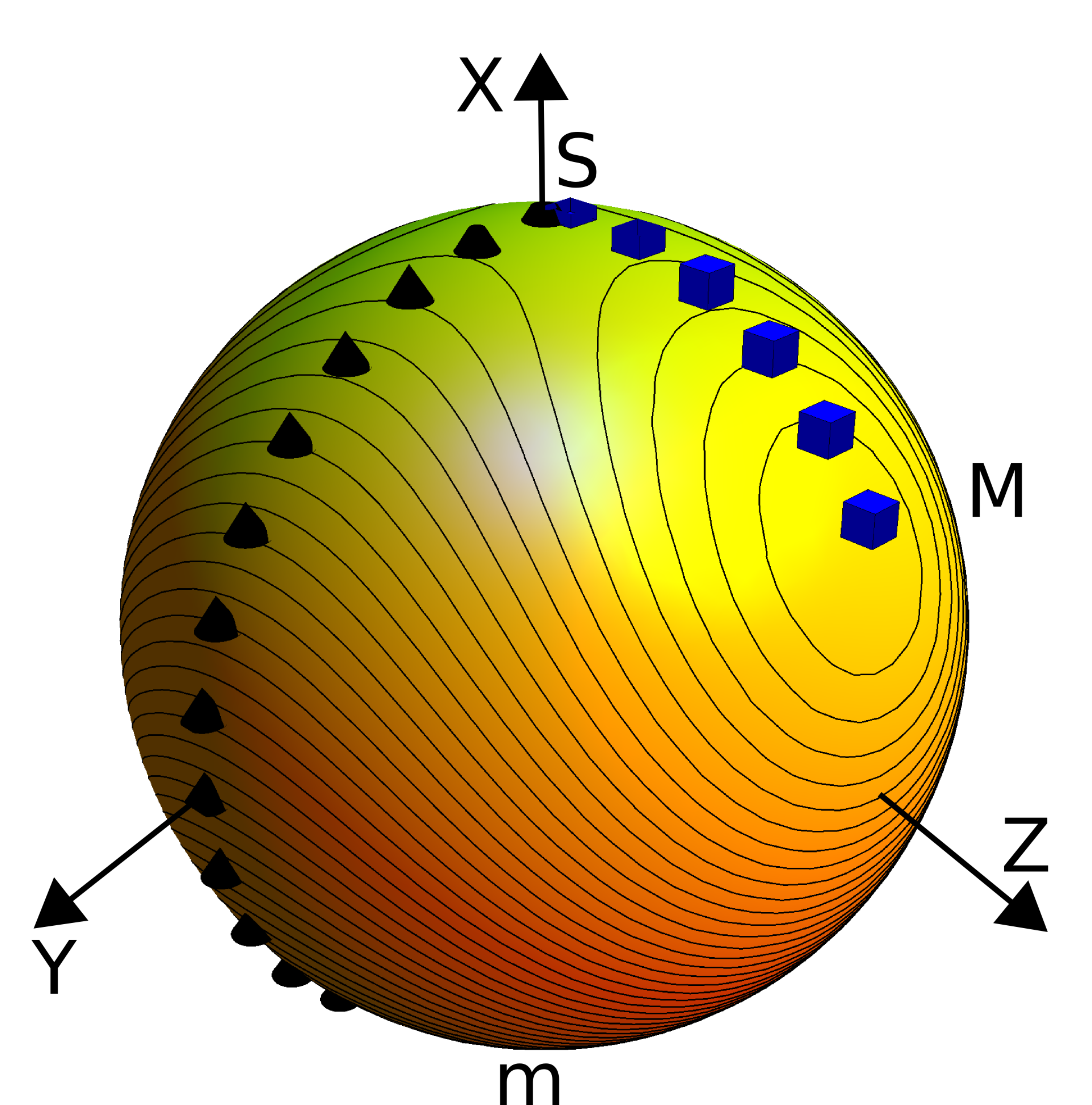}}    
    \put(15,65){ \textbf I)}
    \end{overpic}
    \end{minipage}
    \caption{(color online). Numeric scaled magnetization in the quasienergy eigenstates 
$\expval{J_{x}/j}_{\alpha}$ (filled circles),
the time-averaged expectation value $\overline{\expval{J_{x}/j}}$ (solid, dashed lines), and the semiclassical result $\overline{X}$ (open triangles, squares), see main text.
Inset I) depicts the paths on the 
Bloch sphere to perform the measurement protocol.
    The parameters are $j=40$, $\kappa=0.2$, $p=0.1$, and $T=1$.}
    \label{Fig3}
\end{figure}

We now focus on the description of scaled magnetization (SM) in the quasienergy eigenstates $\expval{J_{x}/j}_{\alpha}=\bra{\Phi_{\alpha}(0)}J_{x}/j\ket{\Phi_{\alpha}(0)}$,  where $\ket{\Phi_{\alpha}(0)}$ is the Floquet mode with quasienergy $E_{\alpha}$.
The filled circles in Fig.~\ref{Fig3} depict the SM as a function of the mean
 quasienergies
$\langle\hat{H}_{\text{E}}\rangle_{\alpha}=E_{\alpha}$.
In the regime $\kappa>p$, the magnetization exhibits a cusp at the critical quasienergy $E_{S}=jp$,

which is a signature of the CQS.
A similar behavior in the SM has been found in undriven LMG-type and Dicke-type models~\cite{Caprio08,BrandesESQPT,PPF11E}.
%%%%
%%%%
%%%%
%%%%
%%%%%%%%%%%%%
\paragraph{Measurement protocol.---}
%%%%%%%%%%%
To resolve the cusp in the 
SM that arises in Fig.~\ref{Fig3}, we propose a measurement protocol which relies on the calculation of time averaged expectation values and is experimentally accessible~\cite{Oberthaler,Chaudhury1,Chaudhury2}.
To initialize the measurement, we consider a spin coherent state 
$\ket{\Psi(0)}=\ket{\gamma(0)}=\sum_{\alpha}a_{\alpha}\ket{\Phi_{\alpha}(0)}$, centered at $\gamma(0)=u(0)+\mathrm{i}v(0)$ on the Bloch sphere~\cite{NoriSpinSqueez}.
The stroboscopic evolution after $l$ kicks is given by $\ket{\Psi(l)}=\hat{\mathcal{F}}^{l}\ket{\Psi(0)}$, such that
the time-averaged density operator after $K$ periods reads $\overline{\rho}=(K+1)^{-1}\sum^{K}_{l=0}\ket{\Psi(l)}\bra{\Psi(l)}$. 
Correspondingly, the time-averaged expectation value of an observable $\hat{A}$ reads
$\overline{\expval{A}}=\rm{tr}(\overline{\rho}\hat{A})$.

In our measurement protocol, the semiclassical trajectories defined by $jE_{G}(\gamma,\gamma^{\ast})= E_{\alpha}$, where $E_{G}$ is the QEL of Eq.~\eqref{QELKicked}, play a fundamental role. 
For unfolded quasienergies below the critical quasienergy $E_{S}=jp$, the trajectories are connected, 
however, when the system crosses $E_{S}$, they become disconnected. 
Therefore, initial states centered along the path joining the saddle $(S)$ with the minimum $(m)$ in the inset I) of Fig.~\ref{Fig3}, enable us to calculate the SM for quasienergies $E_{\alpha}<E_{S}$ [solid, black curve in Fig.~\ref{Fig3}]. 
Correspondingly, the path joining the saddle $(S)$ with the maximum $(M)$ leads to the reconstruction of the SM for $E_{\alpha}>E_{S}$ [dashed, blue curve in Fig.~\ref{Fig3}]. 
These initial states are characterized by a low participation ratio $P^{-1}_{r}=\left(\sum_{\alpha}|a_{\alpha}|^{4}\right)^{-1}$, which implies a high localization over the basis of Floquet modes~\cite{ScharfHomoclinic}. 

To perform the measurement protocol, we must first prepare the system in a desired state $\ket{\Psi(0)}=\ket{\gamma(0)}$ localized around a Floquet mode with quasienergy $E_{\alpha}$, such that $j 
E_{G}[\gamma(0),\gamma^{\ast}(0)]= E_{\alpha}$.  
Next, we let the system evolve for a time long compared with the recurrence time $\tau_R\sim j$, 
dictated by the mean level spacing of the quasienergy spectrum~\cite{Haake1}. 
Measurements of the stroboscopic evolution during $K$ periods allow for the construction of the average 
$\overline{\expval{J_{x}/j}}\approx\sum_{\alpha}|a_{\alpha}|^{2}\expval{J_{x}/j}_{\alpha}\approx\expval{J_{x}/j}_{\alpha}$. 
In addition, as the Floquet mode with quasienergy $E_{\alpha}$ plays a dominant role in the dynamics, the average state $\overline{\rho}$ has a well-defined mean quasienergy $E_{\alpha}=\overline{\expval{H_{E}}}$, where $\hat{H}_{E}$ is given in Eq.~\eqref{EffectiveHamiltonianKicked}. 

The solid and dashed curves in Fig.~\ref{Fig3} depict the 
result of the quantum measurement protocol for $K=700$. 
In addition, the open symbols depict the time averages of the classical magnetization $\overline{X}=(K+1)^{-1}\sum_{l=0}^{K}X(l)$, where
$X(l)=[1-\gamma(l)\gamma^{\ast}(l)]/[1+\gamma(l)\gamma^{\ast}(l)]$. 
The point $\gamma(l)$ is the classical evolution of the initial condition $\gamma(0)$ after $l$ periods~\cite{Haake1,Haake2}. 
Both the quantum as well as the classical result exhibit a cusp at the unfolded
critical quasienergy $E_{S}$, and the measurement protocol is a reasonable approximation to the SM in quasienergy eigenstates.
%

%%%%%%%%%%%%%%%%%%%%
\paragraph{Conclusion.---}
%%%%%%%%%%%%%%%%%%%%

We have shown that signatures analogous to ESQPTs can be found in driven systems. In the driven case the CQS arise due to the dynamical instabilities in the system. 
Since precursors of the CQS appear already at small system sizes, an experimental verification of our results could be carried out by means of setups which realize the kicked top with $j=3$ through 
the hyperfine structure of driven Cesium atoms~\cite{Chaudhury1,Chaudhury2}. 

We expect our findings to generalize to a class of driven systems where an effective semiclassical description is possible, such as the driven LMG~\cite{Engelhardt2013} or Dicke model~\cite{Bastidas2012}.
Points to be addressed in the future are the classification of CQS by parity~\cite{Puebla2013}, the stability of CQS subject to dissipation and their occurrence in strongly driven systems.
%%%%%%%%%%%%%%%%%%%%%%%%%%%%%%%%%%%%%%%%%%%%%%%%%%%%%%%%%%%%%%%%%%%%%%%%%%%%%%%%%%%%%%%%%%%%%%%%%%%%%%%%%%%%%%%%%%%%%%%%%%%
%%%%%%%%%%%%%%%%%%%%%%%%%%%%%%%%%%%%%%%%%%%%%%%%%%%%%%%%%%%%%%%%%%%%%%%%%%%%%%%%%%%%%%%%%%%%%%%%%%%%%%%%%%%%%%%%%%%%%%%%%%%
\paragraph{Acknowledgments.---}
The authors gratefully acknowledge discussions with C. Emary, G. Engelhardt, A. Lazarides, C. Nietner,
and C. Viviescas, and financial
support by the DFG via GRK 1558 (M.V.), grants BRA 1528/7, BRA 
1528/8, SFB 910 (V.M.B., T.B.), the Spanish Ministerio de 
Ciencia e Innovaci\'on (Grants No. FIS2011-28738-C02-01) and Junta de 
Andaluc\'ia 
(Grants No. FQM160) (P.P.-F.). 

%%%%%%%%%%%%%%%%%%%%%%%%%%%%%%%%%%%%%%%%%%%%%%%%%%%%%%%%%%%%%%%%%%%%%%%%%%%%%%%%%%%%%%%%%%%%%%%%%%%%%%%%%%%%%%%%%%%%%%%%%%%
%%%%%%%%%%%%%%%%%%%%%%%%%%%%%%%%%%%%%%%%%%%%%%%%%%%%%%%%%%%%%%%%%%%%%%%%%%%%%%%%%%%%%%%%%%%%%%%%%%%%%%%%%%%%%%%%%%%%%%%%%%%
%%%%%%%%%%%%%%%%%%%%%%%%%%%%%%%%%%%%%%%%%%%%%%%%%%%%%%%%%%%%%%%%%%%%%%%%%%%%%%%%%%%%%%%%%%%%%%%%%%%%%%%%%%%%%%%%%%%%%%%%%%%
%%%%%%%%%%%%%%%%%%%%%%%%%%%%%%%%%%%%%%%%%%%%%%%%%%%%%%%%%%%%%%%%%%%%%%%%%%%%%%%%%%%%%%%%%%%%%%%%%%%%%%%%%%%%%%%%%%%%%%%%%%%
\vspace{-0.7cm}
%%%%%

%%%%%%%%%%%%%%%%%%%%%%%%%%%%%%%%%%%%%%%%%%%%%%%%%%%%%%%%%%%%%%%%%%%%%%%%%%%%%%%%%%%%%%%%%%%%%%%%%%%%%%%%%%%%%%%%%%%%%%%%

%%%%%%%%%%%%%%%
\begin{widetext}
%%%%%%%%%%%%%%%%
 
\section*{Supplementary information for ``Quantum criticality and dynamical instability in the kicked top''}
Equations in the main paper are denoted by Eq.~[*]. 
\section{Calculation of the density of quasienergy states}
%%%%%%%%
%%%%%%%%
Given a system described by a time periodic Hamiltonian $\hat{H}(t)=\hat{H}(t+T)$, one can define the density of quasienergy states (DOQS) as 
$\rho(\varepsilon)=M^{-1}\sum_{\alpha}\delta(\varepsilon-\varepsilon_{\alpha})$, where $\varepsilon_{\alpha}$ are the quasienergies and we have assumed a Hilbert space of dimension $M$ \cite{HaakeBuchSI}. 
Now let us consider a new representation of the DOQS, as follows
%%%
\begin{equation}
      \label{DOQSDef}
            \rho(\varepsilon)=\frac{1}{2\pi M}\sum_{\alpha}\sum_{n=-\infty}^{\infty} e^{-\mathrm{i}n\varepsilon_{\alpha}T}e^{\mathrm{i}n\varepsilon T} =  \frac{1}{2\pi M}\sum_{n=-\infty}^{\infty} t_{n}e^{\mathrm{i}n\varepsilon T}=\frac{1}{2\pi}+\frac{1}{2\pi M}{\rm Re}\left[\sum_{n=1}^{\infty}t_{n}e^{\mathrm{i} n \varepsilon T}\right]
       ,
\end{equation}
%%%
where $t_{n}=\sum_{\alpha}e^{-\mathrm{i}n\varepsilon_{\alpha}T}=\text{tr}\hat{\mathcal{F}}^{n}$ are traces of powers of the Floquet operator $\mathcal{F}$ \cite{HaakeBuchSI}. 
Without loss of generality, we will choose $T=1$ in the following. 

Now we focus on the calculation of the DOQS of the kicked top Hamiltonian~Eq.~[1] (see main text) in the limit $j\gg 1$, which has a Hilbert space of dimension $M=2j+1$.
One can calculate the trace in the basis of spin coherent states $|\gamma\rangle$ 
%%%
\begin{equation}
      \label{SCSTrace}
            t_{n}=\text{tr}\hat{\mathcal{F}}^{n}=\frac{2j+1}{\pi}\int \frac{d^{2}\gamma}{(1+\gamma\gamma^{\ast})^2}\langle \gamma|\hat{\mathcal{F}}^{n}|\gamma \rangle=\frac{2j+1}{\pi}\int \frac{d^{2}\gamma}{(1+\gamma\gamma^{\ast})^2}e^{-\mathrm{i} n jE_{G}(\gamma,\gamma^{\ast})}
       ,
\end{equation}
%%%
where we have used the definition $\hat{\mathcal{F}}=e^{-\mathrm{i}\hat{H}_{E}}$ of the effective Hamiltonian~Eq.~[4] in the main paper. 
The quasienergy landscape (QEL) in local coordinates reads
%%%
\begin{equation}
      \label{QELStereoCoor}
            E_{G}(\gamma,\gamma^{\ast})=\frac{1}{j}\langle \gamma|\hat{H}_{E}|\gamma \rangle=\frac{\kappa}{2}Z^2+\frac{\kappa p Z}{2}\left[X\cot\left(\frac{\kappa Z}{2}\right)-Y \right]
      ,
\end{equation}
%%% 
where we use stereographic coordinates~\cite{HaakeBuchSI}
%%%
\begin{equation}
      \label{StereoCoor}
            (X,Y,Z)= \left(\frac{1-\gamma\gamma^{\ast}}{1+\gamma\gamma^{\ast}},\frac{2{\rm Im}\gamma}{1+\gamma\gamma^{\ast}},\frac{-2{\rm Re}\gamma}{1+\gamma\gamma^{\ast}}\right)
      . 
\end{equation}
%%% 
Motivated by the form of the integral representation of the trace $t_{n}$ in Eq.~\eqref{SCSTrace}, we use the stationary phase approximation formula for $n>0$
%%%
\begin{equation}
      \label{StationaryPhase}
            \int \frac{d^{2}\gamma}{(1+\gamma\gamma^{\ast})^2}e^{-\mathrm{i} n jE_{G}(\gamma,\gamma^{\ast})}=\sum_{c\in \mathcal{C}} \frac{2\pi (1+\gamma_{c}\gamma^{\ast}_{c})^{-2}}{nj\sqrt{|\det\left[\text{HE}_{G}(\gamma,\gamma^{\ast})\right]|_{\gamma=\gamma_{c}}}}e^{\mathrm{i}\beta_{c}\pi/4}e^{-\mathrm{i} n jE_{G}(\gamma_{c},\gamma_{c}^{\ast})}
      ,
\end{equation}
%%%
where $\text{HE}_{G}(\gamma,\gamma^{\ast})$ is the Hessian matrix of the QEL of Eq.~\eqref{QELStereoCoor}, and $\gamma_{c}\in\mathcal{C}$, where $\mathcal{C}$ is the set of critical points that satisfy the conditions $\frac{\partial E_{G}}{\partial \gamma}|_{\gamma=\gamma_{c}}=\frac{\partial E_{G}}{\partial \gamma^{\ast}}|_{\gamma^{\ast}=\gamma^{\ast}_{c}}=0$. The index $\beta_{c}$ is the difference in the number of positive and negative eigenvalues of the Hessian $\text{HE}_{G}(\gamma_c,\gamma^{\ast}_c)$. Therefore, $\beta_{M}=2$ at the maxima and  $\beta_{m}=-2$ at the minima. Similarly
$\beta_{S}=0$ for a saddle point. 

Now we have all the ingredients to evaluate the series in Eq. \eqref{DOQSDef}
%%%
\begin{equation}
      \label{Series}
            \sum_{n=1}^{\infty}t_{n}e^{\mathrm{i} n \varepsilon}= (2j+1)\sum_{c\in \mathcal{C}} \frac{2 (1+\gamma_{c}\gamma^{\ast}_{c})^{-2}}{j\sqrt{|\det\left[\text{HE}_{G}(\gamma,\gamma^{\ast})\right]|_{\gamma=\gamma_{c}}}}e^{\mathrm{i}\beta_{c}\pi/4}
	    {\rm Li}_{1}\left[e^{\mathrm{i}(\varepsilon-E_{c})}\right]  
      ,
\end{equation}
%%%
where $E_{c}=jE_{G}(\gamma_{c},\gamma_{c}^{\ast})$ are the unfolded critical quasienergies, and ${\rm Li}_{1}\left[e^{\mathrm{i}(\varepsilon-E_{c})}\right]=\sum_{n=1}^{\infty}\frac{e^{\mathrm{i} n (\varepsilon- E_{c})}}{n}$ is given in terms of the polylogarithm~\cite{AbramowitzSI}.
This recovers Eq.~[8] in the main manuscript.

%%%%%%%%%%%%%%%%%%%

%%%%%%%%%%%%%%%%%%%

%%%%%%%%%%%%%%%% 
\end{widetext}
%%%%%%%%%%%%%%%%

\end{document}